\documentclass[12pt,a4,epsf]{article}
\usepackage{amsmath,amssymb,amsfonts,picins}
\usepackage[dvips]{graphicx}

\usepackage[bookmarks=true,colorlinks,citecolor=blue,pdftitle={Coupling Constant Dependence of the Kaluza--Klein Spectrum in Five-Dimensional SQCD on S^1},pdfauthor={Yukiko Ohtake, Yuya Wakabayashi}]{hyperref}

\newcommand{\expect}[1]{\langle#1\rangle}
\newcommand{\abs}[1]{\left\lvert#1\right\rvert}
\DeclareMathOperator*{\Tr}{Tr}
\DeclareMathOperator{\Real}{Re}
\DeclareMathOperator{\Image}{Im}
\DeclareMathOperator*{\Arg}{Arg}

\begin{document}

\begin{titlepage}

\begin{flushright}
{\tt hep-th/0511055}\\
OCHA-PP-251\\
November, 2005
\end{flushright}

\vspace{5mm}

\begin{center}
{\Large \bf
Coupling Constant Dependence%
\\of the Kaluza--Klein Spectrum%
\\in Five-Dimensional SQCD on $S^1$}

\vspace{2.5cm}
{\large Yukiko Ohtake}\\
{\it Department of Physics, Ochanomizu University, Tokyo 112-8610, Japan}

\vspace{2.5mm}
{\large Yuya Wakabayashi}\\
{\it Department of Physics, Rikkyo University, Tokyo 171-8501, Japan}
\end{center}

\vspace{2.5mm}
\begin{abstract}
We investigate the Kaluza--Klein (KK) spectrum of $\mathcal{N}\!=\!1$ supersymmetric gauge theory compactified on a circle. We concentrate on a model with gauge group $SU(2)$ and four massless matter fields in the fundamental representation. We derive the exact mass formula of KK modes by using Seiberg--Witten theory. From the mass formula and the D3-brane probe realization, we determine the spectrum of KK modes of matter fields and gauge fields. As a result, we find that the lightest KK state of gauge fields is stable for all the vacuum moduli space, while the lightest KK state of matter fields decays easier than other KK states in a region of the moduli space. The region becomes small as we decrease the five-dimensional gauge coupling constant $g_5$, and vanishes as we take the limit $g_5\to 0$. This result continuously connects the known KK spectrum in the weak  coupling limit and that in the strong coupling limit.
\end{abstract}

\end{titlepage}

\section{Introduction}
Recently, various four-dimensional models embedded in higher dimensional spacetime have been greatly investigated \cite{ExDim}. Their spectra generally include Kaluza--Klein (KK) states, which carry nonzero momentum in extra-dimensions. As an example, let us consider a circle compactification of five-dimensional gauge theory with a massless matter field. The effective four-dimensional theory has a tower of KK modes of the gauge field, say $A_\mu^{(n)}$, and that of the matter field, say $\psi^{(n)}$, where $n$ runs over the integers and $\mu$ runs over $0,1,2,3$. The states $A_\mu^{(n)}$ and $\psi^{(n)}$ carry fifth dimensional momentum, $n/R$, where $R$ is the compactification radius. Then $A_\mu^{(n)}$ and $\psi^{(n)}$ with nonzero $n$ are KK states. In this paper, we investigate the stability of KK states.

A state $A$ is kinematically unstable and decays into states $B$ and $C$ when the charges are conserved in the process and their masses satisfy the inequality $M(A)\geq M(B)+M(C)$. In the case of the five-dimensional model above, masses of $A_{\mu}^{(n)}$ and $\psi^{(n)}$ are classically $\abs{n}/R$. Thus for $n\!\!>\!\!1$, $A_{\mu}^{(n)}$ can decay to $A_{\mu}^{(n-1)}$ and $A_{\mu}^{(1)}$ and $\psi^{(n)}$ can decay to $A_{\mu}^{(n-1)}$ and $\psi^{(1)}$. Similar decay processes occur for $n\!\!<\!\!1$. From these, we see that the stable KK states are $A_{\mu}^{(\pm 1)}$ and $\psi^{(\pm 1)}$. This result is supported by perturbative analysis in \cite{ST}, where the inequality of masses in various models are evaluated under the one-loop correction. On the other hand, nonperturbative behavior of the KK spectrum was found in the strong coupling limit of a supersymmetric extension of the model \cite{Ohtake3}. In \cite{Ohtake3}, the KK spectrum of a circle compactification of the five-dimensional $\mathcal{N}\!=\!2$ supersymmetric model with gauge group $SU(2)$ and $N_f\!=\!5,6,7$ massless fundamental matter fields was studied in the strong coupling limit. The four-dimensional effective theory possesses $\mathcal{N}\!=\!2$ supersymmetry, so that the exact mass formula is derived by using Seiberg--Witten theory \cite{SW}. In addition, the theory has the D3-brane probe realization \cite{BDS}, where the inequality is diagammatically evaluated by using string junctions \cite{BF1}. By using these techniques, it was shown that $A_{\mu}^{(n)}$ can decay to $A_{\mu}^{(n-1)}$ and $A_{\mu}^{(1)}$ similar to the perturbative result, while $\psi^{(n-1)}$ decays easier than $\psi^{(n)}$ in a certain region of vacuum moduli space.

Now we know that the perturbative KK spectrum is different from the spectrum in the strong coupling limit. How does the spectrum change as the five-dimensional coupling constant $g_5$ varies from 0 to $\infty$? To answer the question, we shall generalize the analysis in \cite{Ohtake3} to the case with finite $g_5$. In section 2,  we derive the exact mass formula for finite $g_5$, using Seiberg--Witten theory. From the mass formula and the D3-brane probe realization, we determine the stability of KK modes in section 3. As a result, we show that the nonperturbative behavior also appears in the case of the finite coupling constant. As we decrease $g_5$, the region where the nonperturbative behavior appears becomes small. As we take the limit $g_5\to 0$, the region disappears and the perturbative spectrum is reproduced.

\section{Seiberg--Witten solution}
\subsection{Seiberg--Witten curve}
Five-dimensional $\mathcal{N}\!=\!1$ supersymmetric gauge theory compactified on a circle is effectively described by four-dimensional $\mathcal{N}\!=\!2$ supersymmetric gauge theory. It includes an adjoint complex Higgs scalar field $\phi$ as a superpartner of the gauge field, and has a vacuum moduli space parametrized by the vacuum expectation value of $\phi$. The low energy effective Lagrangian and the mass formula are derived from a Seiberg--Witten curve \cite{SW}.

For the theory with gauge group $SU(2)$ and $N_f$ matter fields in the fundamental representation, which we refer to as quarks, the Seiberg--Witten curve is written as \cite{MNW}
\begin{equation}
y^2 = x^3+f(u)x+g(u), \label{eqn:SW0}
\end{equation}
\begin{align}
f(u) &= \sum_{i=0}^4 a_i u^i,&
g(u) &=\sum_{i=0}^6 b_i u^i,
\end{align}
where $u \!=\!\expect{\Tr\phi^2}$, a gauge invariant moduli parameter, which takes a value on $CP^1$. Constants $a_i$ and $b_i$ depend on the parameters of the theory such as the five-dimensional coupling constant $g_5$, the compactification radius $R$, and masses of the matter fields $m_i$ ($i\!=\!1,\ldots, N_f$). In this subsection, we shall determine these constants. For simplicity, let us assume $N_f\!=\!4$ and $m_i\!=\!0$ for all $i$.

The zero points of the discriminant of \eqref{eqn:SW0},
\begin{equation}
\Delta(u) = 4f(u)^3+27g(u)^2,
\end{equation}
are determined from global symmetry \cite{HO}. The correspondence between global symmetry and zeros of $\Delta(u)$ is listed in \cite{Kodaira} and \cite{dWHIZ}. Since our model has the flavor symmetry $SO(8)$, we obtain that $a_0$ and $b_0$ are to be vanished and $\Delta(u)$ is proportional to $u^6$ \cite{Kodaira}. In addition, the symmetry is extended to broken affine $SO(8)$ \cite{YY}. This is because the states carry not only $SO(8)$ charges but also the KK charge $n$. In other words, an $SO(8)$ multiplet has copies labeled by an integer $n$. This is just the structure of affine $SO(8)$ multiplets. The affine symmetry is broken at the scale $1/R$ because the copies have different masses. The broken affine symmetry requires two additional zero points \cite{dWHIZ}. From these restrictions, we conclude that the curve of our model is
\begin{equation}
\begin{split}
y^2
&= x^3+\left\{a_2+u\left(a_3-\frac{3u}{L^4}\right)\right\}u^2x \\
&\quad +\frac{L^6}{216}\left(a_3-\frac{6u}{L^4}\right)\left\{a_3^2+\frac{24a_3u}{L^4}+\frac{36}{L^4}\left(a_2-\frac{2u^2}{L^4}\right)\right\}u^3.
\label{eqn:SW1}
\end{split}
\end{equation}
To simplify the curve, we scale and shift the variables as  
\begin{align}
y &\to \frac{1}{24\sqrt{3}}y,&
x &\to \frac{1}{12}\left\{x-\frac{L^2}{3}\left(a_3-\frac{u}{L^4}\right)u\right\},&
u &\to \frac{1}{6}u, 
\end{align}
and set $a_3=b/L^2$ and $a_2=(3c^2-b^2)/12$. Then we obtain a simple form,
\begin{equation}
y^2 = x^3+\left(\frac{u}{L^2}-b\right)ux^2+c^2u^2x. \label{eqn:SW2}
\end{equation}
Then the discriminant is given by
\begin{equation}
\Delta(u) = c^4u^6\left\{u-(b+2c)L^2\right\}\left\{u-(b-2c)L^2\right\}.\label{eqn:disc}
\end{equation}

Next we consider the constants $b$, $c$ and $L$. They are functions of $g_5$ and $R$. In the following, we derive explicit forms of the functions by matching the curve \eqref{eqn:SW2} with that of two limits. We choose $R/g_5^2$ and $R$ as a set of independent variables, and complexify $R/g_5^2$ to $\tau=4\pi iR/g_5^2+\theta/2\pi$, the bare coupling constant of four-dimensional effective theory.

Firstly, we consider a limit $R\to 0$ with fixed $\tau$. In this limit,
the theory is reduced to four-dimensional $\mathcal{N}\!=\!2$ $SU(2)$
gauge theory with $N_f\!=\!4$ massless quarks, and the coupling constant
$\tau$. Its Seiberg--Witten curve is known as \cite{SW} 
\begin{equation}
y^2 = x^3-\frac{1}{4}g_2(\tau)u^2x-\frac{1}{4}g_3(\tau)u^3, \label{eqn:SWref}
\end{equation}
where $g_2(\tau)$ and $g_4(\tau)$ are the Eisenstein series:
\begin{align}
g_2(\tau) &= \frac{60}{\pi^4}\sum_{(m,n)\in Z^2_{\neq 0}}\frac{1}{(m+n\tau)^4},&
g_4(\tau) &= \frac{140}{\pi^6}\sum_{(m,n)\in Z^2_{\neq 0}}\frac{1}{(m+n\tau)^6}.
\end{align}
The mass dimensions of $y$, $x$ and $u$ in \eqref{eqn:SWref} are 3, 2 and 2, respectively. Setting the same in \eqref{eqn:SW2}, we find that the mass dimensions of $b$ and $c$ are 0 and that of $L$ is 1. Hence $b$ and $c$ depend only on $\tau$ and then $L$ is written as $1/R$ times a function of $\tau$, say $f(\tau)$. The function $f(\tau)$ is removed from the curve by scaling parameters as $u\to f(\tau)u$, $b\to b/f(\tau)$ and $c\to c/f(\tau)$. Altogether rearranging the curve \eqref{eqn:SW2} yields the following equation:
\begin{equation}
y^2 = x^3+\left(R^2u-b\right)ux^2+c^2u^2x \label{eqn:SW3}.
\end{equation}
Now we take the limit $R\to 0$ in \eqref{eqn:SW3} and compare with \eqref{eqn:SWref}. As we shift $x$ as $x\to x-bu/3$ in \eqref{eqn:SWref}, we see that
\begin{align}
b &= -3e_i(\tau) \label{eqn:b}\\
c^2 &= 3e_i^2(\tau)+e_1(\tau)e_2(\tau)+e_2(\tau)e_3(\tau)+e_3(\tau)e_1(\tau).
\label{eqn:c}
\end{align}
Here $e_i(\tau)$ ($i\!=\!1,2,3$) are the solutions of the equation $x^3-g_2(\tau)x/4-g_3(\tau)/4=0$. They are written as
\begin{equation}
\begin{aligned}
e_1(\tau) &= \frac{1}{3}(-\theta_1^4(\tau)+2\theta_3^4(\tau)),\\
e_2(\tau) &= \frac{1}{3}(-\theta_1^4(\tau)-\theta_3^4(\tau)),\\
e_3(\tau) &= \frac{1}{3}(2\theta_1^4(\tau)-\theta_3^4(\tau)), \label{eqn:es}
\end{aligned}
\end{equation}

where $\theta_1$ and $\theta_3$ are
\begin{align}
\theta_1(\tau) &= \sum_{n\in Z}e^{i\pi\tau (n+1/2)^2},&
\theta_3(\tau) &= \sum_{n\in Z}e^{i\pi\tau n^2}. \label{eqn:theta}
\end{align}

Secondly, we consider the limit $g_5\to\infty$ with zero $\theta$-angle and fixed $R$. In this limit, flavor symmetry $SO(8)$ is enhanced to $E_5$ \cite{Seiberg} and the curve should be \cite{MNW,YY}
\begin{equation}
y^2 = x^3+(R^2u-4)ux^2+4u^2x. \label{eqn:SWref2}
\end{equation}
Now we take the limit in the curve \eqref{eqn:SW3}. From \eqref{eqn:es}, \eqref{eqn:theta}, and the relations $e_1(\tau)=e_2(-1/\tau)/\tau^2$, $e_2(\tau)=e_1(-1/\tau)/\tau^2$ and $e_3(\tau)=e_3(-1/\tau)/\tau^2$, we see that $e_1(\tau)\sim -1/3\tau^2$, $e_2(\tau)\sim 2/3\tau^2$ and $e_3(\tau)\sim -1/3\tau^2$ in the limit. Thus $b\sim 1/\tau^2$ and $c^2\sim 0$ when $i$ in \eqref{eqn:b} and \eqref{eqn:c} is $1$ or $3$, while $b\sim 2/\tau^2$ and $c^2\sim 1/\tau^4$ for $i\!=\!2$. In the former case, we cannot make \eqref{eqn:SW3} coincide with \eqref{eqn:SWref2}. In the latter case, we make it by scaling parameters as $x\to x/4\tau^4$, $y\to y/8\tau^6$ and $u\to u/2\tau^2$. Then we choose $i\!=\!2$ in \eqref{eqn:b} and \eqref{eqn:c}. Thus we have
\begin{align}
b &= \theta_1^4(\tau) +\theta_3^4(\tau), \label{eqn:bf}\\
c &= \theta_1^2(\tau)\theta_3^2(\tau). \label{eqn:cf}
\end{align}

In summary, the Seiberg--Witten curve of our model is \eqref{eqn:SW3}, where $b$ and $c$ are given by \eqref{eqn:bf} and \eqref{eqn:cf}. Its discriminant is \eqref{eqn:disc} with $L\!=\!1/R$. The zero points of $\Delta (u)$ are located at $u\!=\!0$ and $(b\pm 2c)/R^2$. It is known that extra massless states appear at each zero point \cite{SW}. For simplicity, we set $R\!=\!1$ in the following.

\subsection{Mass formula}
From the Seiberg--Witten curve \eqref{eqn:SW3}, we shall derive the mass formula of stable states called BPS states. For this purpose, we derive the periods of the curve,
\begin{equation}
\Pi(u)=
\begin{pmatrix}
\omega_D(u) \\
\omega(u)
\end{pmatrix}
=
\begin{pmatrix}
\oint_\beta\frac{dx}{y} \\
\oint_\alpha\frac{dx}{y}
\end{pmatrix}, \label{eqn:pi}
\end{equation}
where $\alpha$ and $\beta$ are the homology cycles on the torus given by \eqref{eqn:SW3} with a fixed $u$.

The periods $\Pi(u)$ are determined from the Picard--Fuchs equation,
\begin{equation}
\left\{\frac{d^2}{du^2}
+\frac{3u^2-4bu+b^2-4c^2}{u(u-b-2c)(u-b+2c)}\frac{d}{du}
+\frac{4u^2-2bu-b^2+4c^2}{4u^2(u-b-2c)(u-b+2c)}\right\}\Pi(u)=0.\label{eqn:period0}
\end{equation}
As we set $\Pi(u)=u^{-1/2}k(w)$ with $w=-\left\{u-(b+2c)\right\}/4c$, the equation becomes
\begin{equation}
\frac{d^2 k}{dw^2}
+\frac{1-2w}{w(1-w)}\frac{dk}{dw}
-\frac{1/4}{w(1-w)}k
=0.
\end{equation}
This is the standard hypergeometric equation of $\alpha\!\!=\!\!\beta\!\!=\!\!1/2$ and $\gamma\!=\!1$. It has two independent solutions,
\begin{align}
K(w) &= \int_{-i\infty}^{i\infty}\frac{ds}{2\pi i}
\left\{\Gamma(-s)\Gamma(\frac{1}{2}+s)\right\}^2(1-w)^s, \label{eqn:K}\\
K'(w) &= \int_{-i\infty}^{i\infty}\frac{ds}{2\pi i}
\left\{\Gamma(-s)\Gamma(\frac{1}{2}+s)\right\}^2w^s. \label{eqn:Kd}
\end{align}
Series expansions of $K(w)$ and $K'(w)$ for $\abs{w}<1$, $\abs{1-w}<1$ and $\abs{1/w}<1$ are easily derived. For instance, the expansions for $\abs{w}\!<\!1$ are
\begin{align}
K(w) &= \pi\sum_{n=0}^{\infty}
\left\{\frac{\Gamma(\frac{1}{2}+n)}{n!}\right\}^2w^n, \label{eqn:EK}\\
K'(w) &= -\sum_{n=0}^{\infty}\left\{\frac{\Gamma(\frac{1}{2}+n)}{n!}\right\}^2w^n
\left\{\log w+4\sum_{r=0}^{n-1}\left(\frac{1}{2r+1}-\frac{1}{2r+2}\right)-2\log 4\right\}.\label{eqn:EKd}
\end{align}
The periods are given by linear combinations of the functions,
\begin{equation}
\Pi(u) = u^{-\frac{1}{2}}
\begin{pmatrix}
c_1 & c_2 \\
c_3 & c_4
\end{pmatrix}
\begin{pmatrix}
K(w) \\
K'(w)
\end{pmatrix}.
\end{equation}
Coefficients $c_1, \ldots, c_4$ are determined by direct calculation of elliptic integrals \eqref{eqn:pi} for $\abs{w}\!<\!1$ and comparing the result with the expansions \eqref{eqn:EK} and \eqref{eqn:EKd}. Indeed we have
\begin{equation}
\begin{pmatrix}
c_1 & c_2 \\
c_3 & c_4
\end{pmatrix}
=\frac{1}{\sqrt{c}\,\pi^2}
\begin{pmatrix}
2\pi & \phantom{m} & 0 \\
-\pi-2i\log 4 & \phantom{m} & -i\pi
\end{pmatrix}.
\end{equation}

The periods undergo monodromy around zeros of $\Delta(u)$. The monodromy matrices acting on $\Pi(u)$ around $u\!=\!0$, $b-2c$ and $b+2c$ are
\begin{align}
M_0 &= \begin{pmatrix}-1&0\\0&-1\end{pmatrix}, &
M_{-}& = \begin{pmatrix}3&4\\-1&-1\end{pmatrix}, &
M_{+}& = \begin{pmatrix}1&0\\-1&1\end{pmatrix},
\end{align}
respectively. From these matrices, we can determine what kind of states becomes massless at the zeros; when the monodromy matrix around a first order zero is
\begin{equation}
M_{(p,q)} =
\begin{pmatrix}
1-pq & \phantom{m} & p^2 \\
-q^2 & \phantom{m} & 1+pq
\end{pmatrix},
\end{equation}
a state $(p,q)$, whose electric and magnetic charges of unbroken $U(1)$ gauge symmetry are $p$ and $q$ respectively, becomes massless. Thus we see that $(2,-1)$ becomes massless at $u\!=\!b-2c$ and $(0,1)$ at $u\!=\!b+2c$. Moreover, six states become massless simultaneously at $u\!=\!0$, the sixth order zero. Since $M_0$ is expressed as $M_{(1,0)}^4M_{(2,-1)}M_{(0,1)}$, the massless states are $(2,-1)$, $(0,1)$ and four $(1,0)$.

Now we can write down the mass formula. It is described by the integration of the periods by $u$ \cite{SW}. Each bound of the integrals is decided in order to reproduce the extra massless states at the zero points correctly. Then the mass formula of a state with the electric and magnetic charges $(p,q)$ and the charges of broken affine symmetry related to the singularities at $u\!=\!b\pm2c$, say $n_1$ and $n_2$, is given by
\begin{align}
M_{(p,q,n_1,n_2)}(u) &= \abs{Z_{(p,q,n_1,n_2)}(u)}, \label{eqn:mass}\\
Z_{(p,q,n_1,n_2)}(u) &= pa(u)-qa_D(u)+n_1s_1+n_2s_2, \label{eqn:central}
\end{align}
where
\begin{align}
a_D(u) &= \int_0^{u}\!\!\omega_D(u')du',\\
a(u) &= \int_0^{u}\!\!\omega(u')du',\\
s_1 &= \int_0^{b+2c}\!\!\omega_D(u)du=8\pi\arcsin\sqrt{\frac{b+2c}{4c}},\\
s_2 &= -\int_0^{b-2c}\!\!\left\{2\omega(u)+\omega_D(u)\right\}du
   =8\pi\arcsin\sqrt{\frac{b-2c}{4c}}.
\end{align}
\piccaptionoutside
\piccaption{\footnotesize Three brunch cuts.\label{setup}}
\parpic[r]{\includegraphics[bb=50 50 220 201,width=200pt,clip]{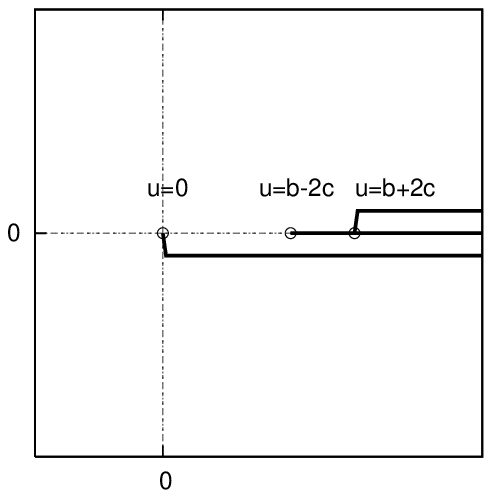}}\picskip{14}
In the following,  we evaluate the values of $a$ and $a_D$ by means of the numerical integration on {\it Mathmatica}. For simplicity, we assume that the $\theta$-angle is zero. Then $b$ and $c$ are real, $c\!>\!0$ and $b\!\!>\!\!2c$. Hence the zeros of $\Delta(u)$, $u\!=\!0$, $b-2c$ and $b+2c$, are aligned from the left on the real axis of the $u$-plane. We set the branch cuts of $a_D$ and $a$ from these zeros to $\infty$ along the real axis as depicted in Fig.\ref{setup}. When we cross each cut from the lower half $u$-plane, \\$^t\!(a_D(u),a(u),s_1,s_2)$ is changed by the matrices
\begin{align}\hspace{-1em}
\widetilde{M}_0 &= \left(\begin{smallmatrix}-1&0&\phantom{0}0\phantom{0}&\phantom{0}0\phantom{0}\\0&-1&0&0\\0&0&1&0\\0&0&0&1\end{smallmatrix}\right), &
\widetilde{M}_{-} &= \left(\begin{smallmatrix}-1&-4&-2&\phantom{0}0\phantom{0}\\1&3&1&0\\0&0&1&0\\0&0&0&1\end{smallmatrix}\right), &
\widetilde{M}_{+} &= \left(\begin{smallmatrix}\phantom{0}1\phantom{0}&\phantom{0}0\phantom{0}&\phantom{0}0\phantom{0}&0\\1&1&0&-1\\0&0&1&0\\0&0&0&1\end{smallmatrix}\right)
\end{align}
for the cut from $u\!=\!0$, $b-2c$ and $b+2c$, respectively. To conserve \eqref{eqn:central}, we should change the charges $^t\!(p,q,n_1,n_2)$ simultaneously. The matrices acting on $^t\!(p,q,n_1,n_2)$ are given by
\begin{align}\hspace{-1em}
{K}_0 &= \left(\begin{smallmatrix}-1&0&\phantom{0}0\phantom{0}&\phantom{0}0\phantom{0}\\0&-1&0&0\\0&0&1&0\\0&0&0&1\end{smallmatrix}\right),&
{K}_{-} &= \left(\begin{smallmatrix}-1&-4&\phantom{0}0\phantom{0}&\phantom{0}0\phantom{0}\\1&3&0&0\\-1&-2&1&0\\0&0&0&1\end{smallmatrix}\right),&
{K}_{+} &= \left(\begin{smallmatrix}1&\phantom{0}0\phantom{0}&\phantom{0}0\phantom{0}&\phantom{0}0\phantom{0}\\1&1&0&0\\0&0&1&0\\-1&0&0&1\end{smallmatrix}\right).
\label{eqn:mono}
\end{align}

\section{Stability of KK modes}
Four-dimensional theory described by a Seiberg--Witten curve \eqref{eqn:SW0} appears in type IIB string theory as the world volume theory of a D3-brane probe in a 7-brane background \cite{BDS}. For our model, we employ the background constructed from six 7-branes $[1,0]^4[2,-1][1,0]$ at $z\!=\!0$, a 7-brane $[2,-1]$ at $z\!=\!b-2c$, and a 7-brane $[0,1]$ at $z\!=\!b+2c$. Here $z$ is a complex coordinate of the space transverse to the 7-branes, $[1,0]$ denotes a D7-brane, and $[p,q]$ denotes an $SL(2,Z)$-dual 7-brane \cite{Schwarz}. The metric on the $z$-plane is given by \cite{cosmic}
\begin{equation}
ds^2 = \Image\widetilde{\tau}(z)\abs{\frac{da(z)}{dz}dz}^2,
\end{equation}
where $\widetilde{\tau}(z)\!=\!da_D(z)/da(z)$. In this background, the world volume theory of a D3-brane probe located at $z=u$ is our model with the moduli parameter $u$. States of the model correspond to strings ending on the D3-brane. Therefore, the spectrum of states corresponds to the spectrum of strings which can end on the D3-brane probe. Then to find the spectrum of KK modes, we study the spectrum of corresponding strings.

In IIB string theory, there appear a fundamental string, $(1,0)$, and its $SL(2,Z)$-dual strings, $(p,q)$. A string $(p,q)$ has its ends on a D3-brane or a 7-brane with the same charges, $[p,q]$. In addition, strings can merge each other and make a string junction \cite{Schwarz}. Thus there are three kinds of strings ending on the D3-brane probe: a string connecting a 7-brane and the D3-brane, a string with the both ends on the D3-brane, and a string emanating from a string junction and ending on the D3-brane. Some strings are related to string junctions transitionally, due to the string creation at 7-branes \cite{transit}. In any case, a string $(p,q)$ ending on the D3-brane is detected as a state with the electric charge $p$ and the magnetic charge $q$. Especially, a string $(1,0)$ connecting $[1,0]$ at $z=0$ and the D3-brane corresponds to a quark, and a string $(1,0)$ with the both ends on the D3-brane corresponds to a gauge field. The winding number of a string around the 7-branes is equivalent to the KK charge $n$ \cite{MOY}.

To be stable, a string stretches along a geodesic that minimizes the string mass. The mass of a string $(p,q)$ along a curve $C$ is given by
\begin{equation}
\int_CT_{(p,q)}ds = \int\abs{dZ_{(p,q,0,0)}},
\end{equation}
where $T_{(p,q)}=\abs{p-q\widetilde{\tau}}/\sqrt{\Image\widetilde{\tau}}$, the tension of $(p,q)$ \cite{Schwarz}. In order to minimize this, points on a geodesic emanating from $z\!=\!z_0$ satisfy the equation
\begin{equation}
\Arg\left\{Z_{(p,q,0,0)}(z)-Z_{(p,q,0,0)}(z_0)\right\} = \phi. \label{eqn:geodesic}
\end{equation}
Here $\phi$ is a constant between $0$ and $2\pi$. It takes the same value for geodesics of the strings in a stable junction \cite{BF1}. Then, the mass of a string junction constructed from $n_1(2,-1)$ from $z\!=\!b-2c$, $n_2(0,1)$ from $z\!=\!b+2c$, $(p-2n_1,q+n_1-n_2)$ from $z\!=\!0$ and an outgoing string $(p,q)$ coincides with \eqref{eqn:mass} when the outgoing string ends on the D3-brane at $z\!=\!u$. Thus the junction corresponds to a state with the charges $(p,q,n_1,n_2)$. Note that the geodesic equation of the outgoing string $(p,q)$ is rewritten as
\begin{equation}
\Arg\{Z_{(p,q,n_1,n_2)}(z))\} = \phi. \label{eqn:geod}
\end{equation}
Hereinafter, we refer to the string obeying \eqref{eqn:geod} as $(p,q,n_1,n_2)$. When a string $(p,q,n_1,n_2)$ crosses branch cuts, it undergoes monodromy described by the matrices \eqref{eqn:mono}.

Let us notice that $\phi$ parameterizes the direction of geodesics. As
$\phi$ varies, the geodesic \eqref{eqn:geodesic} moves around the point
$z\!=\!z_0$ and sweeps some region in the $z$-plane. When the region
includes a point $z\!=\!u$, the string can end on the D3-brane probe at
$z\!=\!u$ and is detected as a stable state. Therefore, the region in
the $z$-plane through which the geodesic of a string passes corresponds
to the region in the moduli $u$-plane where the corresponding state is
stable \cite{BF1}. Then in the following, we seek the regions of KK modes of quarks $\psi^{(n)}$, unbroken $U(1)$ gauge fields $A_\mu^{(n)}$ and $W$-bosons $W_\mu^{(n)}$ by evaluating \eqref{eqn:geodesic} and \eqref{eqn:geod} for corresponding strings. In the following, we assume $b-2c\!=\!4$ and $b+2c\!=\!6$ unless we explicitly state otherwise.

\begin{figure}[t]
\begin{center}
\includegraphics[bb=50 50 274 258,width=360pt,clip]{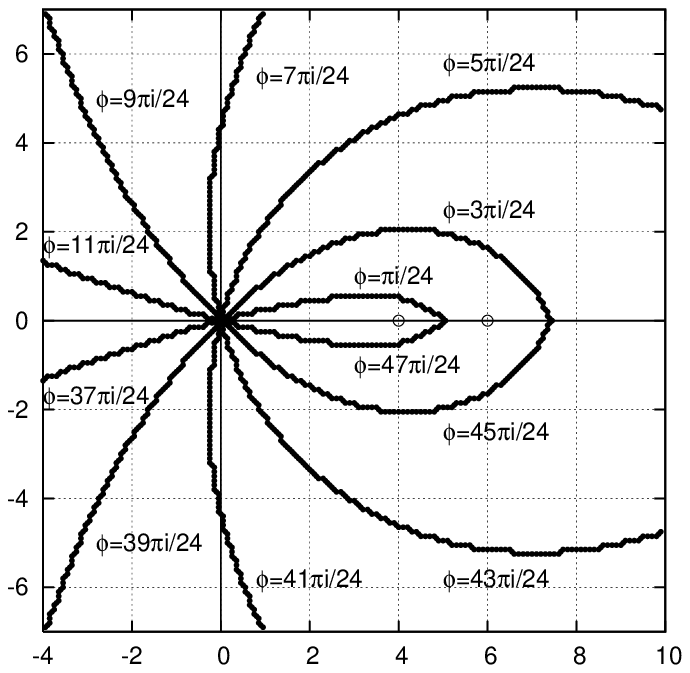}
\end{center}
\caption{\footnotesize The string which corresponds to a quark KK zero mode $\psi^{(0)}$ sweeps out whole of the $z$-plane. This result means that $\psi^{(0)}$ exists stably for any value of the moduli parameter $u$.}
\label{GroundStates}
\end{figure}%
\paragraph*{Quarks}
We now consider the cases of quarks. First of all we consider $\psi^{(0)}$. It corresponds to a string $(1,0)$ emanating from $z\!=\!0$, that is, $(1,0,0,0)$. The geodesics for various $\phi$ are evaluated as depicted in Fig.\ref{GroundStates}. From the figure, we see that the string sweeps all the $z$-plane. Thus we can conclude that $\psi^{(0)}$ is stable for all the moduli $u$-plane.

\begin{figure}[t]
\begin{center}
\begin{tabular}{ccc}
{\scriptsize A}\hspace{-1em}\includegraphics[bb=50 50 220 210,width=140pt,clip]{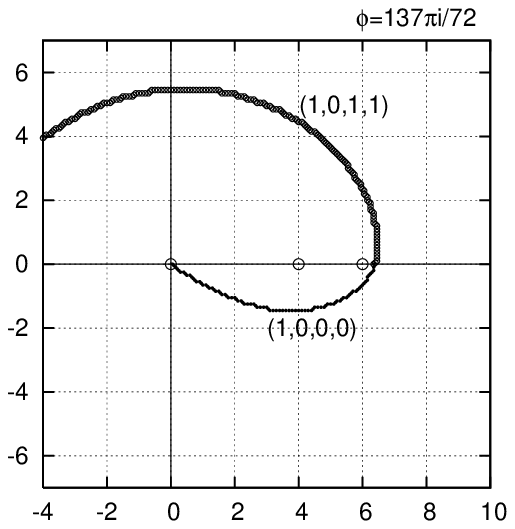}&\hspace{-1.5em}
{\scriptsize B}\hspace{-1em}\includegraphics[bb=50 50 220 210,width=140pt,clip]{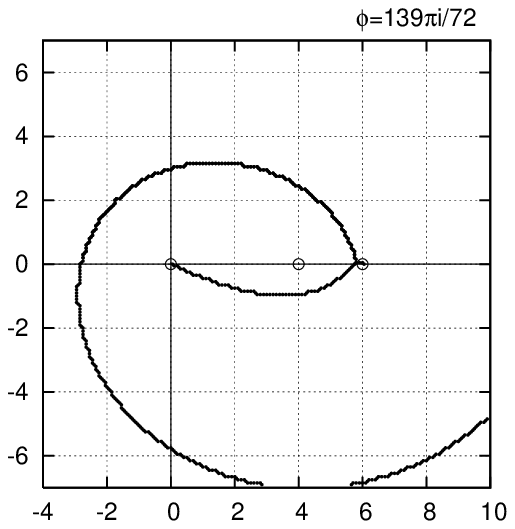}&\hspace{-1.5em}
{\scriptsize C}\hspace{-1em}\includegraphics[bb=50 50 220 210,width=140pt,clip]{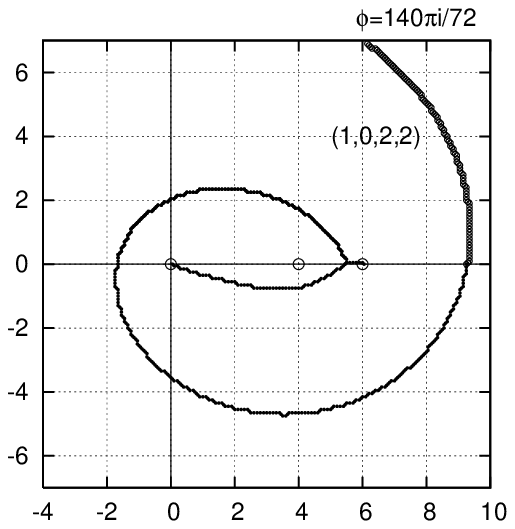}\\
{\scriptsize D}\hspace{-1em}\includegraphics[bb=50 50 220 210,width=140pt,clip]{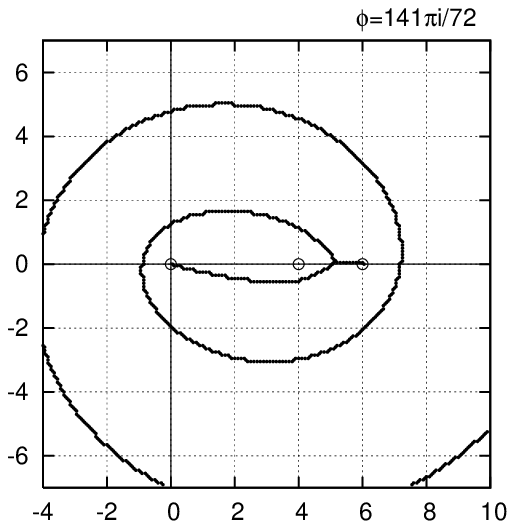}&\hspace{-1.5em}
{\scriptsize E}\hspace{-1em}\includegraphics[bb=50 50 220 210,width=140pt,clip]{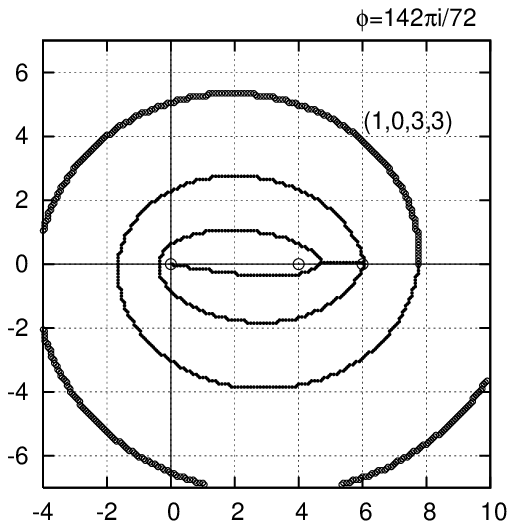}&\hspace{-1.5em}
{\scriptsize F}\hspace{-1em}\includegraphics[bb=50 50 220 210,width=140pt,clip]{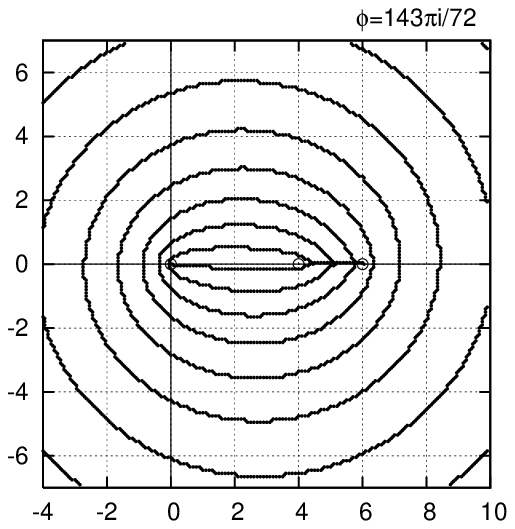}\\
\end{tabular}
\end{center}
\caption{\footnotesize The string geodesics corresponding to $\psi^{(n)}$. The string $(1,0,n,n)$ corresponds to the quark $\psi^{(n)}$. Here, we show the strings up to $n\!=\!8$. In Fig.\ref{FirstPlots}A--F, some string junctions are constructed. For details, see Fig.\ref{closeup}.}
\label{FirstPlots}
\end{figure}
\begin{figure}[t]
\begin{center}
\includegraphics[bb=50 45 410 302,width=\textwidth,clip]{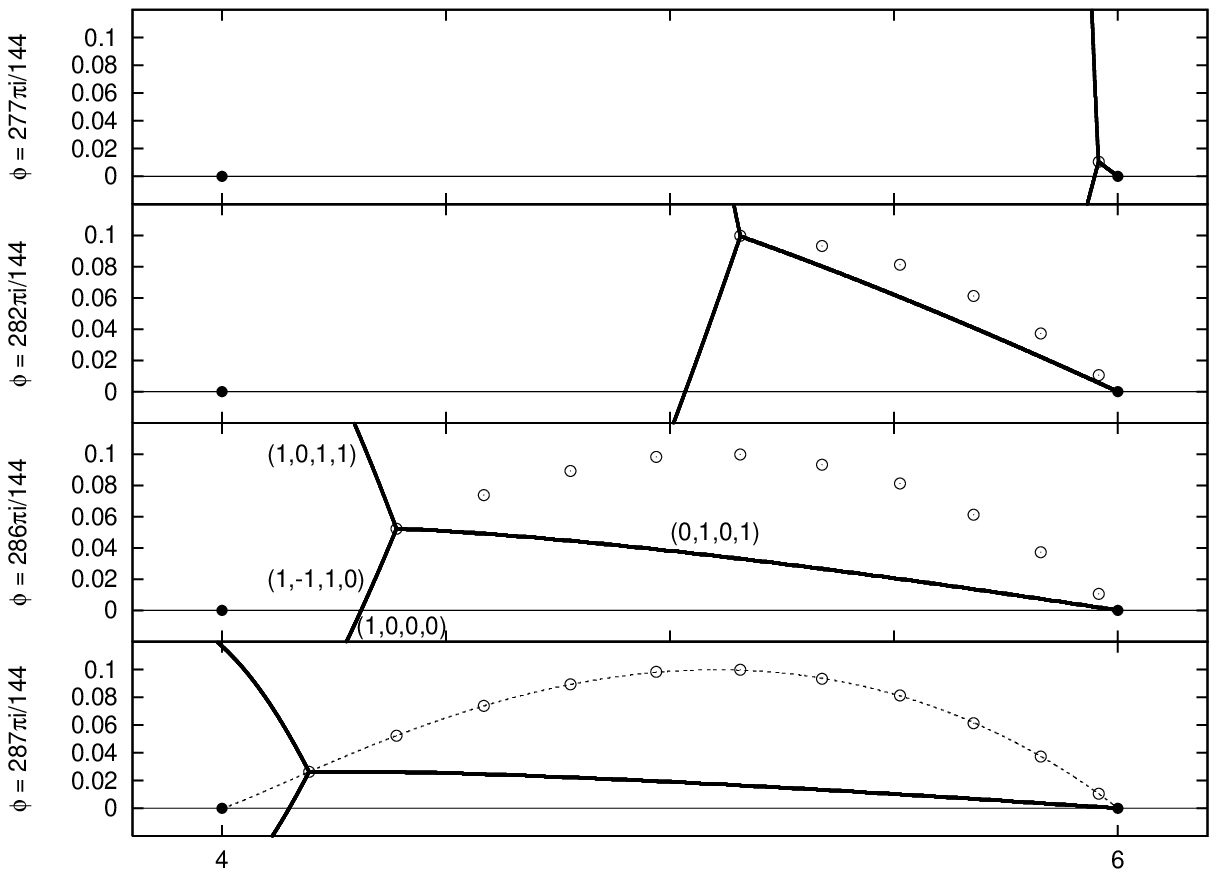}
\end{center}
\caption{\footnotesize The enlarged illustrations of a string junction with the outgoing string $(1,0,1,1)$. We illustrate the point where the strings merge by a blob $\circ$.  Connecting these blobs, we get a curve stretched from the 7-brane at $z\!=\!b+2c\,(=\!6)$ to the one at $z\!=\!b-2c\,(=\!4)$. We call this curve $C_1$, the curve of marginal stability for $\psi^{(1)}$.}
\label{closeup}
\end{figure}

%
Secondly we consider $\psi^{(1)}$. It corresponds to a string $(1,0,0,0)$ going around the 7-branes once anti-clockwisely. The string crosses the branch cuts in $\Real z\!>\!b+2c$. Then the charges are changed to $K_+K_-K_0{}^t\!(1,0,0,0)={}^t\!(1,0,1,1)$ as depicted in Fig.\ref{FirstPlots}A. Thus $\psi^{(1)}$ corresponds to a string $(1,0,1,1)$. The string sweeps a region in the first quadrant of the $z$-plane, as we increase $\phi$. Moreover, as depicted in Fig.\ref{FirstPlots}B--C, the string hits the 7-brane $[0,1]$ at $z\!=\!b+2c$ and becomes a string junction. The detailed configuration near the merging point of strings is shown in Fig.\ref{closeup}. The junction is constructed from three strings: $(1,-1,1,0)$ which comes from $(1,0,0,0)$ crossing the cuts between $z\!=\!0$ and $z\!=\!b-2c$, $(0,1,0,1)$ emanating from $[0,1]$ at $z\!=\!b+2c$, and the outgoing string $(1,0,1,1)$. As we increase $\phi$ further, the merging point draws a curve $C_1$ as shown in Fig.\ref{closeup}, and $(1,0,1,1)$ sweeps outside the region $S_1$ surrounded by $C_1$ and the real axis. Thus $\psi^{(1)}$ is stable outside $S_1$ in the moduli $u$-plane and disappears inside it. In addition, when the D3-brane is located on $C_1$, $(1,0,1,1)$ can decay into $(0,1,0,1)$ and $(1,-1,1,0)$. Thus $\psi^{(1)}$ becomes marginally stable when the moduli parameter $u$ is on $C_1$.

\begin{figure}[t]
\begin{center}
\includegraphics[bb=50 50 425 260,width=\textwidth,clip]{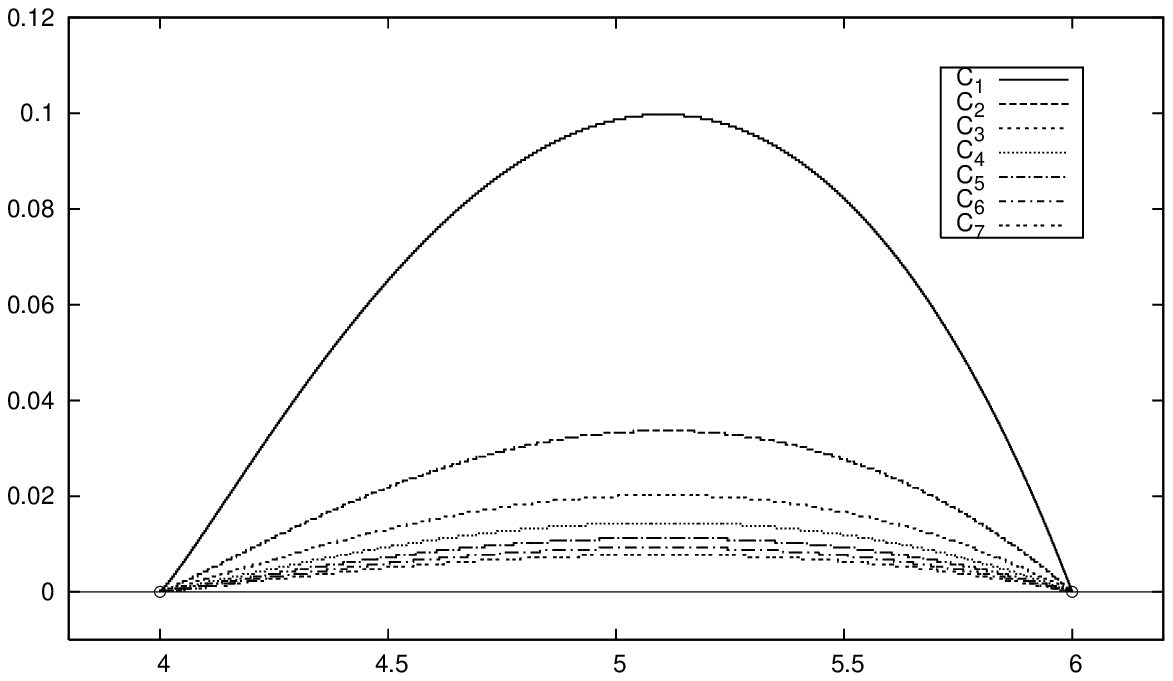}
\end{center}
\caption{\footnotesize The marginal stability curves for $\psi^{(n)}$ ($1\leq n\leq7$). Here we also set $z\pm2c=6,4$.}
\label{CMSs}
\end{figure}

Thirdly we consider $\psi^{(2)}$. It corresponds to $(1,0,2,2)$, which comes from a string $(1,0,1,1)$ going around the 7-branes once more (see Fig.\ref{FirstPlots}C--D). As depicted in Fig.\ref{FirstPlots}E, the string hits the 7-brane at $z\!=\!b+2c$ again and becomes a complicated string junction. As we increase $\phi$, the merging point of the strings draws a curve $C_2$, and $(1,0,2,2)$ sweeps outside the region $S_2$ surrounded by $C_2$ and the real axis. Thus $\psi^{(2)}$ is stable outside the region and disappears inside it. Note that the merging point is apparently under $C_1$. Consequently, $S_2$ is inside $S_1$, as shown in Fig.\ref{CMSs}.

On the basis of these observations, we next consider $\psi^{(n)}$ ($n\!\geq\!3$). In general, $\psi^{(n)}$ corresponds to a string $(1,0,n,n)$, which appears as $(1,0,n\!-\!1,n\!-\!1)$ crosses the branch cuts in $\mbox{Re}z\!>\!b+2c$. The string hits $[0,1]$ at $z\!=\!b+2c$ and an additional string $(0,1,0,1)$ is created. The merging point of the strings draws a curve $C_n$, which connects $z\!=\!b\pm 2c$ in the upper half $u$-plane, as we increase $\phi$. Simultaneously, $(1,0,n,n)$ sweeps outside the region $S_n$ surrounded by $C_n$ and the real axis. Thus $\psi^{(n)}$ appears outside $S_n$ and disappears inside it. The region $S_n$ is inside $S_{n-1}$ as shown in Fig.\ref{CMSs}. Therefore, $\psi^{(n)}$ disappears in numeric order of $n$ as we change the moduli parameter from a value above $C_1$ to the segment $[b-2c,b+2c]$. Similar analysis can be done for $\psi^{(n)}$ with negative $n$. It corresponds to a string $(1,0,0,0)$ going around the 7-branes $\abs{n}$ times clockwisely. One can derive the curve of marginal stability for $\psi^{(n)}$, and obtain the mirror image of that for $\psi^{(-n)}$ with respect to the real axis in the $u$-plane. These results are similar to the quark KK spectrum in the strong coupling limit $g_5\to\infty$ \cite{Ohtake3}.

\begin{figure}[h]
\begin{center}
\includegraphics[bb=60 50 402 196,width=\textwidth]{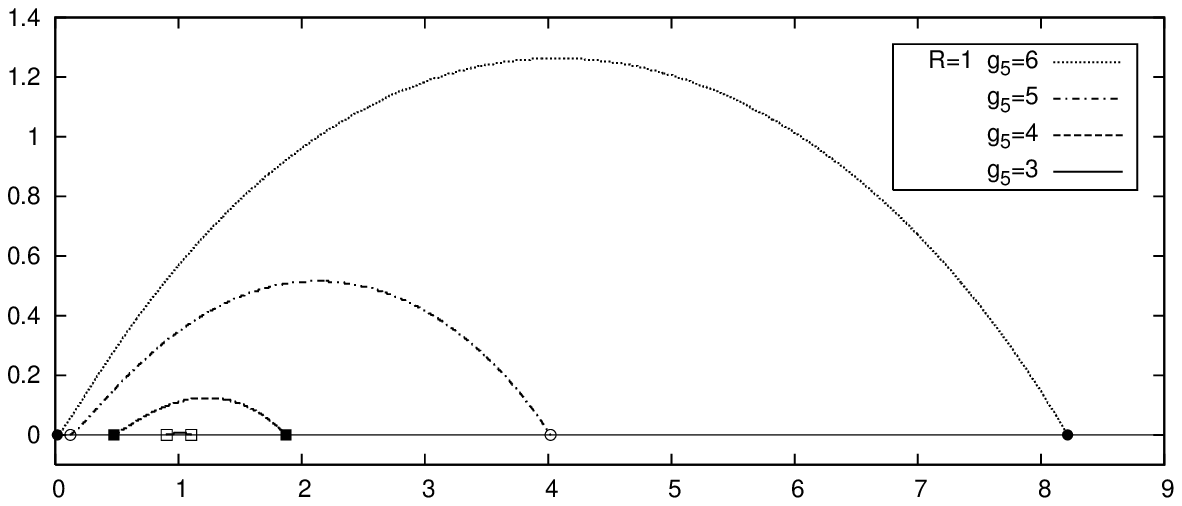}
\end{center}
\caption{\footnotesize The marginal stability curve $C_1$ of $\psi^{(1)}$ for $g_5\!=\!3,4,5,6$. The weaker $g_5$ becomes, the shorter the curve $C_1$ becomes.}
\label{BCflow}
\end{figure}

\piccaptionoutside
\piccaption{\footnotesize The positions of two singularities at $z\!=\!b\pm2c$ depend on $g_5$. On the one hand, they collide in the limit of $g_5\to0$. On the other, the singularity $z\!=\!b-2c$ collides with the one at $z\!=\!0$ in the strong coupling limit, $g_5\to\infty$.\label{ccFig}}
\parpic[r]{\includegraphics[bb=50 45 260 198,width=200pt,clip]{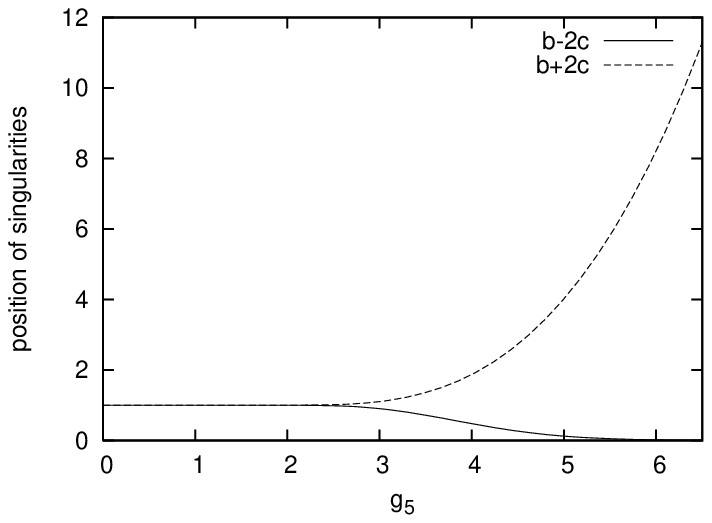}}\picskip{15}
Now we determine the $g_5$ dependence of the quark KK spectrum. From \eqref{eqn:bf} and \eqref{eqn:cf}, we see that the positions of singularities, $z\!=\!b\pm 2c$, depend on $g_5$ as shown in Fig.\ref{ccFig}. The distance of the singularities becomes small as we decrease $g_5$. Then the curves of marginal stability, which connect the two singularities, also become small. As an example, we show the curve of $\psi^{(1)}$ for $g_5\!=\!3,4,5,6$ in Fig.\ref{BCflow}. The curve shrinks to a point in the limit $g_5\to 0$, where two singularities at $z\!=\!b\pm 2c$ collide. Therefore, the nonperturbative jumps of the quark KK spectrum disappear in this limit, as was expected from the perturbative analysis. On the other hand, in the limit $g_5\to\infty$, $z\!=\!b-2c$ coincides with $z\!=\!0$. Then the quark KK spectrum in the strong coupling limit is reproduced.

\paragraph*{$W$-bosons}
Next we are interested in $W_\mu^{(0)}$. It corresponds to a string $(1,0)$ emanating from the D3-brane, going around the 7-branes at $z\!=\!0$, and coming back to the D3-brane. The string crosses the branch cut from $z\!=\!0$ and becomes a string $(-1,0)$ because of the monodromy. Both the geodesic of $(1,0)$ and that of $(-1,0)$ are the same with that of $(1,0,0,0)$, a string corresponding to $\psi^{(0)}$. Therefore, the region where the geodesic for $W_\mu^{(0)}$ sweeps is the same with that for $\psi^{(0)}$. Then the stability of $W_\mu^{(0)}$ is the same with $\psi^{(0)}$; it is stable for all the moduli $u$-plane. In addition, a KK state $W_\mu^{(n)}$ corresponds to a string corresponding to $W_\mu^{(0)}$ which goes around the 7-branes $n$ times anti-clockwisely. The string has the same geodesic for $\psi^{(n)}$. Thus we conclude that $W_\mu^{(n)}$ is stable outside $S_n$ and disappears inside it.

\paragraph*{Unbroken $U(1)$ gauge bosons}
Before concluding this paper, we shall consider unbroken $U(1)$ gauge bosons. We start with $A_{\mu}^{(0)}$. It corresponds to a string $(1,0)$ localized on the D3-brane probe at $z\!=\!u$. Since the string appears for any value of $u$, $A_{\mu}^{(0)}$ is stable for all the $u$-plane.

\begin{figure}[t]
\begin{center}
\begin{tabular}{ccc}
{\scriptsize A}\hspace{-1em}\includegraphics[bb=50 50 220 210,width=140pt,clip]{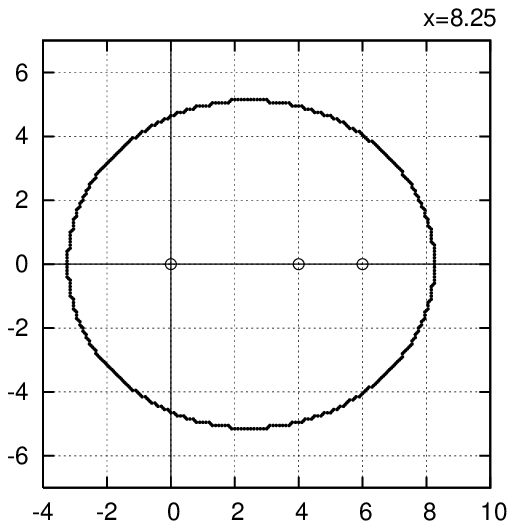}&\hspace{-1.5em}
{\scriptsize B}\hspace{-1em}\includegraphics[bb=50 50 220 210,width=140pt,clip]{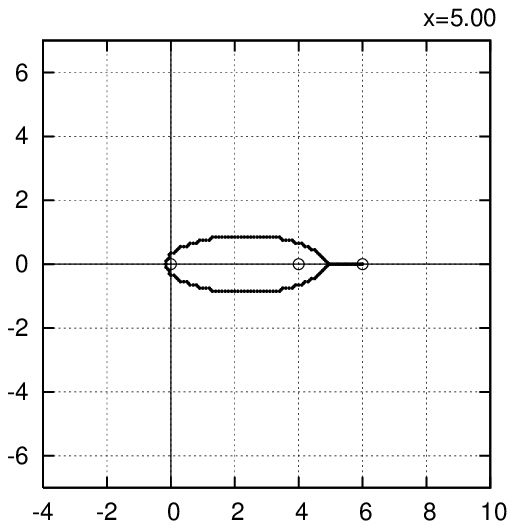}&\hspace{-1.5em}
{\scriptsize C}\hspace{-1em}\includegraphics[bb=50 50 220 210,width=140pt,clip]{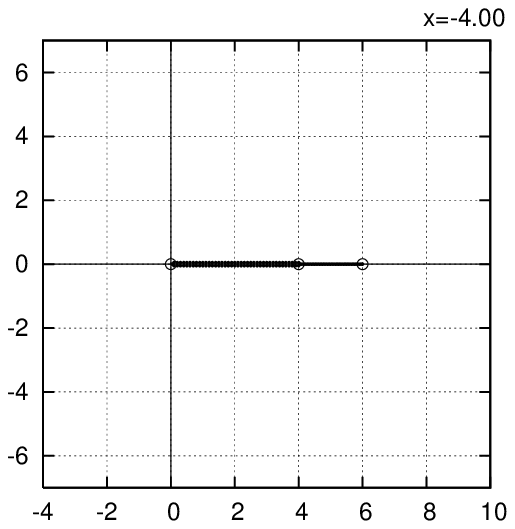}\\
\end{tabular}
\end{center}
\caption{\footnotesize The string geodesics corresponding to the lightest gauge KK state $A_\mu^{(1)}$.}
\label{SecondPlots}
\end{figure}

Next we consider $A_{\mu}^{(1)}$. It corresponds to a string $(1,0)$ emanating from the D3-brane probe, going around all the 7-branes once anti-clockwisely, and coming back to the D3-brane. The geodesic equation is given by 
\begin{equation}
\Arg\{a(z)-a(x)\} = 0,
\end{equation}
where $x$ is a point where the string passes through. The geodesics for various $x$ are loops as shown in Fig.\ref{SecondPlots}. As we decrease $x$ from $\infty$ to $b+2c$, the loop becomes small. When $x\!=\!b+2c$, the loop hits the 7-brane at $z\!=\!b+2c$ and becomes a string junction as depicted in Fig.\ref{SecondPlots}B. As we decrease $x$ from $b+2c$ to $b-2c$, the loop part of the junction collapses to a straight line as depicted in Fig.\ref{SecondPlots}C. In any case, when the D3-brane is located on the loop, the junction cannnot decay and is observed as a stable $A_{\mu}^{(1)}$. Since the loop sweeps all the $z$-plane but the segment between $z\!=\!0$ and $z\!=\!b+2c$, $A_\mu^{(1)}$ is stable for all the $u$-plane but the segment. On the other hand, when the D3-brane is located on the segment, the junction can decay into two parts.  Thus the segment is the marginal stability curve of $A_\mu^{(1)}$.

Finally, we consider $A_\mu^{(n)}$ ($n\!\!>\!\!1$). This corresponds to a string $(1,0)$ going around the 7-branes $n$ times. The string corresponds to $n$ loop strings or loop string junctions derived for $A_\mu^{(1)}$. Thus $A_\mu^{(n)}$ can decay into $n$ $A_\mu^{(1)}$ for any value of $u$. Similary, one can derive that $A_{\mu}^{(-1)}$ is stable for all the $u$-plane but the segment and $A_\mu^{(-n)}$ ($n\!\!>\!\!1$) can decay into $n$ $A_{\mu}^{(-1)}$.

\section*{Acknowledgements}
Y.~O. would like to thank A.~Sugamoto and  T.~Suzuki for useful discussions.

\end{document}